# Stable bright and vortex solitons in photonic crystal fibers with inhomogeneous defocusing nonlinearity

Valery E. Lobanov[1], Olga V. Borovkova[1], Yaroslav V. Kartashov[1,2], Boris A. Malomed,[3] and Lluis Torner[1]

[1]ICFO-Institut de Ciencies Fotoniques, and Universitat Politecnica de Catalunya, 08860 Castelldefels (Barcelona), Spain
[2]Institute of Spectroscopy, Russian Academy of Sciences, Troitsk, Moscow Region, 142190, Russia
[3]Department of Physical Electronics, School of Electrical Engineering, Tel Aviv University, Tel Aviv 69978, Israel
*Corresponding author: valery.lobanov@icfo.eu

We predict that a photonic crystal fiber whose strands are filled with a defocusing nonlinear medium can support stable bright and also vortex solitons if the strength of the defocusing nonlinearity grows toward the periphery of the fiber. The domains of soliton existence depend on the transverse growth rate of the filling nonlinearity and nonlinearity of the core. Remarkably, solitons exist even when the core material is linear.

Localized nonlinear excitations or optical solitons exist in many nonlinear materials and physical settings [1]. In spatially homogeneous media, focusing and defocusing nonlinearities support spatial bright and dark solitons, respectively. However, the situation may drastically change in the presence of transverse modulation of the refractive index. For example, periodic modulations, in the form of the so-called photonic lattices, affect the strength and even sign of the effective diffraction for the propagating beams, so that, under appropriate conditions, bright gap solitons may emerge even in defocusing media (see recent reviews [2,3] and references therein).

The local nonlinearity of materials can be made spatially inhomogeneous too. The propagation of light in the corresponding nonlinear and mixed linear-nonlinear lattices has drawn much attention (for a recent review, see Ref. [4]). Solitons in such structures may exhibit unusual properties, because the corresponding effective inhomogeneity of the material depends on the intensity of the light beam [5-12]. In contrast to periodic linear lattices or localized linear waveguiding structures, structures with localized or periodic defocusing nonlinearities do not support bright solitons [13].

However, it was recently shown that, in contrast to common expectations, a spatially inhomogeneous defocusing nonlinearity whose strength grows sufficiently fast toward the periphery of a material does support stable bright solitons in all dimensions, $D=1,2,3$ [14,15]. The existence of such solitons is directly related to the fact that the growth of the local nonlinearity coefficient makes the governing evolution equation non-linearizable for decaying soliton tails, in contrast to media with uniform or periodic nonlinearities, where the presence of the decaying tails places soliton into the semi-infinite spectral gap of the linearized system, in which defocusing nonlinearities cannot support localization. The settings introduced in Refs. [14,15] require the local strength of the defocusing nonlinearity to grow toward the periphery faster than $r^D$.

In this Letter we show that formation of solitons is also possible in inhomogeneous defocusing nonlinearity landscape where the local nonlinearity varies in a step-like fashion, even when narrow defocusing areas alternate with linear domains in the transverse plane. We thus predict that stable bright two-dimensional (2D) fundamental and vortex solitons exist in photonic-crystal fibers (PCFs) whose strands are selectively filled by refractive-index-matched materials with a suitable defocusing nonlinearity. Such setting may be implemented in liquid-infiltrated PCFs, a technology that now well established [16,17]. Inhomogeneous nonlinear landscapes appear also in various settings [4], such as doped photorefractive materials (e.g. $LiNbO_3$) [18].

A necessary ingredient for the formation of solitons is a step-wise growth of the nonlinearity coefficient in the filled strands towards the periphery of the PCF, while the nonlinearity may remain transversally constant inside each hole. Surprisingly, solitons are found to exist even if the material forming the PCF is linear.

We address the propagation of light beams along the $\xi$-axis in a medium with a uniform linear refractive index and a transverse modulation of the Kerr nonlinearity, that is described the nonlinear Schrödinger equation for the dimensionless field amplitude $q$:

$$i\frac{\partial q}{\partial \xi} = -\frac{1}{2}\left(\frac{\partial^2 q}{\partial \eta^2} + \frac{\partial^2 q}{\partial \zeta^2}\right) - \sigma(\eta,\zeta)|q|^2 q. \quad (1)$$

Here the transverse coordinates $\eta, \zeta$ and propagation distance $\xi$ are normalized to the characteristic transverse scale and diffraction length, respectively, and the function $\sigma(\eta,\zeta)$ describes the nonlinearity profile. The holes in the PCF are arranged into a perfectly periodic hexagonal structure [see Fig. 1(a)]. We assume that they are filled with index-matched defocusing materials so that the nonlinearity coefficient in the central hole is $\sigma(\eta=\zeta=0)=-1$, and its value grows toward the PCF periphery as $\sigma(\eta,\zeta)=-\exp(\alpha r_{km})$, where $r_{km}$ is the distance between the center of the hole with indices $k,m$ and the PCF axis, $\eta, \zeta = 0$. The PCF core material may also be nonlinear, with $\sigma = \sigma_c$. We consider only the case of a linear or focusing core material, with $\sigma_c \geq 0$. Here we set the PCF pitch (hole-to-hole separation) to $d = 1.5$, the hole radius to $r_0 = 0.6$, and the rate of the nonlinearity growth to $\alpha = 1.5$, but we have verified that the results remain qualitatively similar for other values of these parameters.

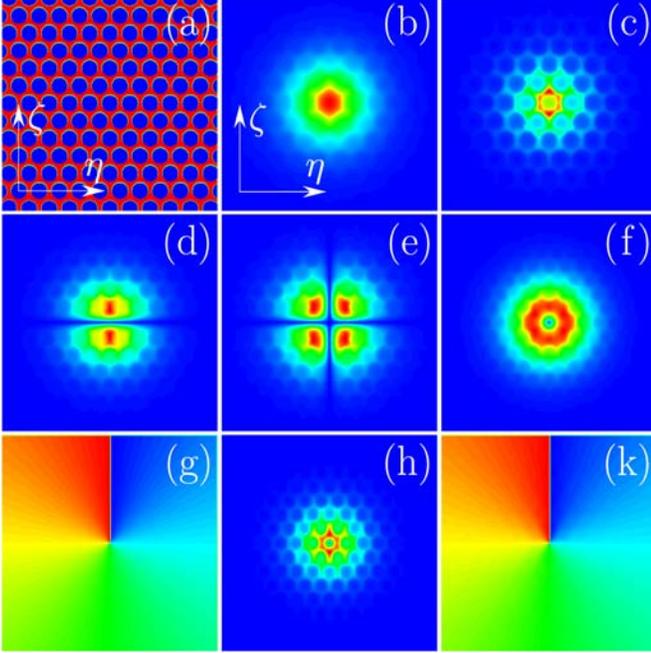

Fig. 1. (a) The cross-section of photonic crystal fiber, where core is shown by red and filled strands are shown by blue. Distribution of the absolute value of the field for fundamental solitons with $b = -1$ (b) and $b = -6$ (c), a dipole soliton with $b = -2$ (d), a quadrupole soliton with $b = -2$ (e), and vortex solitons with topological charge 1 and $b = -1$ (f) or $b = -4$ (h). Panels (g) and (k) show phase distributions for vortex solitons depicted in (f) and (h), respectively. In all the cases, $\sigma_c = 0$.

We search for soliton solutions of Eq. (1) in the form $q(\eta, \zeta, \xi) = w(\eta, \zeta) \exp(ib\xi)$, where $b$ is the propagation constant, while the function $w(\eta, \zeta)$ is real for bright solitons, and complex for vortex modes. In the case of a uniform defocusing nonlinearity ($\alpha = 0$), bright solitons exist only if the nonlinearity of the host PCF is self-focusing, with $\sigma_c > 0$. Such solitons exhibit a ring-like shape because light is expelled from the central hole containing the defocusing medium into the surrounding focusing area. Such solitons are unstable, as they collapse upon the propagation.

For a nonzero growth rate of the defocusing nonlinearity, $\alpha > 0$, different families of bright fundamental, multipole, and vortex solitons exist for $\sigma_c \geq 0$. Illustrative examples are shown in Fig. 1. All such solitons feature a strong small-scale shape modulation, reflecting the underlying PCF structure, and cover multiple holes of the PCF. The modulation is most pronounced at high powers, when light is expelled from the holes into the host material. Note that all solitons in Fig. 1 were obtained for $\sigma_c = 0$, i.e., for a linear host material. Remarkably, even in this case the light field remains localized around the center of PCF. We also found solitons for an algebraic modulation law of the local defocusing nonlinearity in the holes, $\sigma(\eta, \zeta) = -1 - r_{km}^{2+\varepsilon}$, with $\varepsilon > 0$, similar to the results reported for a continuous medium in Ref. [14]. In this case, the slower nonlinearity growth rate results in a weaker soliton localization.

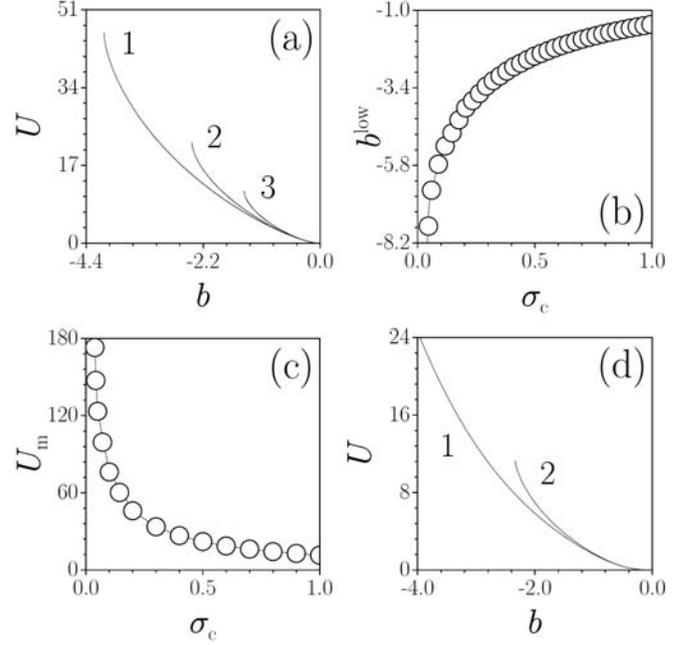

Fig. 2. (a) Energy flow of the fundamental solitons versus propagation constant at $\sigma_c = 0.2$ (curve 1), 0.5 (curve 2), and 1.0 (curve 3). Lower cutoff (b) and the corresponding energy flow $U_m = U(b = b^{low})$ versus the nonlinearity of the host material. (d) Energy flow of the vortex solitons versus propagation constant at $\sigma_c = 0$ (curve 1) and $\sigma_c = 1$ (curve 2).

Solitons exist at negative values of the propagation constant, viz., $b^{low} \leq b < 0$ in the case of fundamental solitons [see Fig. 2(a)], with the energy flow being a monotonically decreasing function of $b$, with $U \to 0$ at $b \to 0$. In contrast, at $b = b^{low}$ the tangential line to the $U(b)$ curve becomes vertical, although $U$ remains finite at $b = b^{low}$. Thus, the energy flow of the fundamental solitons is limited by the value $U_m = U(b = b^{low})$. The cutoff $b^{low}$ depends on the nonlinearity strength $\sigma_c$ of the host material [Fig. 2(b)]. The smaller $\sigma_c$ the broader the existence domain, and $b^{low} \to -\infty$, $U_m \to \infty$ at $\sigma_c \to 0$ [Fig. 2(c)]. Thus, while structures with a focusing host material support fundamental solitons with limited energy flows, there is no limitation on the energy flow in the case of PCF made of a linear material.

Note that at $\sigma_c = 0$ an increase of $U$ gradually leads to a concentration of light in the linear medium, as visible by comparing Figs. 1(b) and 1(c). The soliton width monotonically decreases with a decrease of $b$ (i.e., an increase of $|b|$). The cutoff value is also found to depend on the radius $r_0$ of the PCF holes. When $r_0$ decrease the cutoff $b^{low}$ monotonically increases, while the respective largest energy flow $U_m$ decreases. For all parameters fixed, the cutoff $b^{low}$ is a monotonically decreasing function of the nonlinearity growth rate $\alpha$. To understand the origin of the limit $U_m$, note that a larger part of the energy flow in Fig. 1(c) is carried by the host material. Therefore, self-focusing ($\sigma_c > 0$) leads to collapse at $U = \sigma_c^{-1} U_0$, where the collapse threshold at $\sigma_0 = 1$ may be estimated as the norm of the Townes soliton [1], $U_0 \approx 5.85$. The resulting estimate $U_m \approx 5.85/\sigma_c$ is consistent with Fig. 2(c).

Similar results are obtained for the other types of localized modes, including multipole and vortex solitons. For example, typical dependencies $U(b)$ for single-charged vortex solitons are presented in Fig. 2(d), for cases when $\sigma_c = 0$ and when $\sigma_c > 0$.

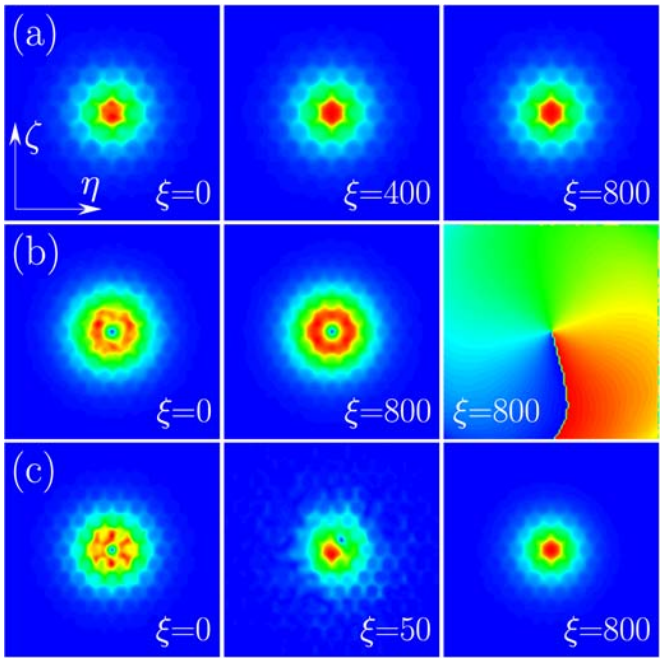

Fig. 3. Stable propagation of perturbed fundamental solitons with $b = -2$ (a) and vortex soliton with $b = -1$ (b), and decay of an unstable vortex with $b = -2$ into a stable fundamental soliton (c). Distributions of the absolute value of the field are shown at different propagation distances. For the stable vortex soliton, the final phase distribution is also shown. In all cases $\sigma_c = 0$.

To check the stability of the stationary soliton families, we studied their propagation up to large distances $\xi \sim 10^4$ in the presence of significant input perturbations to the exact soliton profiles. The fundamental solitons are found to be stable in their entire existence domain. Perturbed inputs reshape and then propagate in a stable way over indefinitely long distances [Fig. 3(a)]. On the other hand, dipole and quadrupole solutions were found to be always unstable. In the course of the propagation, they exhibit fast reshaping into fundamental solitons. Single-charge vortex solitons were also found to be stable inside a finite domain $b^{st} \le b < 0$. An example of the stable propagation of a vortex soliton is shown in Fig. 3(b). Outside the stability domain, i.e., at $b < b^{st}$, they reshape and decay into fundamental solitons [Fig. 3(c)]. Their stability border is $b^{st} \approx -1.48$ at $\sigma_c = 0$, and $b^{st} \approx -1.33$ at $\sigma_c = 1$, i.e., the stability domain shrinks with increase of the strength of the focusing nonlinearity of the host material.

To study the robustness of the solutions against finite-size effects in the infiltrated PFCs, we studied truncated structures where the strength of the defocusing nonlinearity inside the holes grows only for $0 \le r_{km} \le r_{max}$, while at $r_{km} > r_{max}$ it saturates at the level of $\sigma = \sigma(r_{max})$. As expected on physical grounds, input solutions quickly diffract for small values of $r_{max}$. However, already for $r_{max} = 5d$ they retain their shape for more than 200 diffraction lengths without considerable distortion.

Summarizing, we have shown that stable bright two-dimensional and vortex solitons exist in PCFs infiltrated with an inhomogeneous defocusing nonlinearity without any modulation of the linear refractive index, provided that the defocusing nonlinearity grows towards the periphery of the fiber. The central result is thus that bright solitons may be self-sustained in suitable inhomogeneous defocusing nonlinear media without a linear refractive index modulation.

This work was supported by the Ministry of Science and Innovation, Government of Spain, grant FIS2009-09928. Discussions with H. Giessen and M. Vieweg (University of Stuttgart) are gratefully acknowledged.